\documentclass[12pt]{iopart}
\usepackage{graphicx}
\usepackage[sort&compress,comma,square,numbers,longnamesfirst]{natbib}
\usepackage[capitalize]{cleveref}
\usepackage{dcolumn}
\usepackage{bm}
\usepackage{siunitx}

\begin{document}

\title[Surface Enhanced Infrared Absorption and plasmonics]{Surface Enhanced Infrared Absorption mechanism and modification of the plasmonic response}

\author{Tanguy Colleu$^1$, Adam Fekete$^1$\footnote{Present address : Institut für Physik, Humboldt-Universität zu Berlin, Newtonstraße 15, 12489 Berlin, Germany}, Xavier Gonze$^2$, Alexandre~Cloots$^2$, Vincent Liégeois$^3$, Gian-Marco Rignanese$^2$, Luc~Henrard$^1$}

\address{$^1$Solid State Laboratory (LPS), Namur Institute of Structured Matter (NISM), University of Namur, Rue de Bruxelles 61, 5000 Namur, Belgium}
\ead{\mailto{tanguy.colleubanse@unamur.be}, \mailto{luc.henrard@unamur.be}}
\address{$^2$Institute of Condensed Matter and Nanosciences,
Université catholique de Louvain, Rue de l'observatoire 8,
bte L07.03.01, B-1348 Louvain-la-Neuve, Belgium}
\address{$^3$Theoretical Chemistry Laboratory (LCT), Namur Institute of Structured Matter (NISM), University of Namur, Rue de Bruxelles 61, 5000 Namur, Belgium}

\vspace{10pt}
\date{\today}

\begin{abstract}
Surface Enhanced Infrared Absorption (SEIRA) is an experimental method where trace amount of a compound can be detected with high sensibility. This high detection sensibility is the result of the interaction of the molecules with a localized plasmon, usually from a metallic nano-particle. In this study we numerically investigate by discrete dipole approximation the origin of the Fano-like response of the system, including the induced transparency when the plasmon resonance and the molecular vibrational mode coincide. The detailed analysis of the localization of the absorption show that the modification of the absorption cross-section when the molecule is present comes from a change of the plasmonic resonance, not from the direct molecular response which is negligible. This sheds a new light on the SEIRA mechanism. In particular, it demonstrates that the sensibility is associated with the influence of the molecule on the plasmon resonance rather than with the local field enhancement itself. 
\end{abstract}



\maketitle

\section{Introduction}

Vibrational spectroscopies are powerful optical characterization techniques that offer the advantages of being label free and non-destructive. In addition to Raman spectroscopy, infrared (IR) absorption is widely used for direct analysis of molecular functional groups and in situ analysis of surface reactions~\cite{hwang_ultrasensitive_2021,hu_gas_2019,baillieul_surface_2022,ishikawa_cross-polarized_2017,osawa_surface-enhanced_1993,Kozuch_23}. However, its small absorption cross-section, although much larger than that of Raman spectroscopy, limits its utility for large quantities of materials. In contrast, surface-enhanced infrared absorption (SEIRA), based on the coupling with a plasmonic system, increases the sensibility of the detection by a factor of 10\textsuperscript{1} to 10\textsuperscript{7}, enabling optical sensing of trace amounts of molecules~\cite{hu_gas_2019,baillieul_surface_2022,ishikawa_cross-polarized_2017,osawa_surface-enhanced_1993,neubrech_surface-enhanced_2017,dong_nanogapped_2017,Kozuch_23}.

SEIRA necessitates that the plasmonic system resonates at the same frequency as that of the specific molecular vibrational mode under consideration. In practice, gold nanoparticles are frequently utilized due to their high chemical stability and their localized surface plasmon resonance (LSPR) in the IR range. Gold nanorods are particularly popular due to the good control of the synthesis or fabrication processes and the high tunability of their optical resonance, through adjusting their length and radius~\cite{giordano_self-organized_2020,neuman_importance_2015,abb_surface-enhanced_2014,Kozuch_23}. Alternative nanoparticle shapes with sharper edges such as gold bowtie~\cite{dong_nanogapped_2017} or trimers~\cite{mackin_plasmonic_2019} have been considered. Besides gold nanoparticles, graphene is also a substrate of choice due to its tunable IR plasmon via doping~\cite{hu_gas_2019,farmer_ultrasensitive_2016,li_graphene_2014,yan_tunable_2014}.

The interpretation of SEIRA relies on the comparison of the absorbance or transmittance associated with the LSPR with or without the presence of the probed molecules. When both resonances coincide, the optical response diminishes with the presence of the molecules, resulting in a negative differential cross-section and an induced transparency phenomenon. If the resonances are slightly mismatched, a Fano-like behavior is observed~\cite{giordano_self-organized_2020,neuman_importance_2015,dong_nanogapped_2017,li_graphene_2014,yan_tunable_2014,zvagelsky_plasmonic_2021,cerjan_asymmetric_2016,langer_present_2020,neubrech_plasmonic_2013,neubrech_spatial_2014,neubrech_surface-enhanced_2017,wei_ultrasensitive_2019,tanaka_nanostructure-enhanced_2022}. The remarkable sensibility of SEIRA is linked to the fact that the magnitude of the differential cross-section is orders of magnitude larger than that of isolated molecules~\cite{hu_gas_2019,baillieul_surface_2022,ishikawa_cross-polarized_2017,osawa_surface-enhanced_1993,neubrech_surface-enhanced_2017,dong_nanogapped_2017}. This observation is often associated with the enhancement of the local electromagnetic field at the molecule's position~\cite{langer_present_2020,zvagelsky_plasmonic_2021,li_graphene_2014,Kozuch_23}. However, this interpretation does not account for the induced transparency, or the Fano-like shape~\cite{yang_nanomaterial-based_2018,neuman_importance_2015}. The relationship of the SEIRA with the square of the local electric field which follows this first analysis has also never been clearly evidenced. 

In an early investigation~\cite{osawa_surface-enhanced_1993} and in a more recent review~\cite{yang_nanomaterial-based_2018}, it was suggested that the influence of the molecules on the polarizability of the metal nanoparticles at the LSPR is the underlying mechanism for SEIRA. This argument which was based on simulations of the optical response of ellipsoidal metallic nanoparticles coated with a dielectric layer, posits that the plasmonic particles act as an antenna, relaying the signal of the molecule, rather than just exalting the electromagnetic fields.

In this study, we present a comprehensive examination of the SEIRA mechanism through numerical simulations. The manuscript is organized as follows. Our numerical simulations show that the SEIRA effect is associated with a surprisingly high influence of the molecules on the LSPR resonance. They demonstrate that the molecule differential cross-section does not originate from changes of the absorption cross-section of the molecule. Furthermore, our findings corroborate that the scattering cross-section exhibits a similar behavior as the absorption cross-section in SEIRA, despite the fact that molecules themselves do not scatter light~\cite{neuman_importance_2015}.

In Sec.II, we introduce discrete dipole approximation (DDA)~\cite{draine_discrete-dipole_1994,flatau_fast_2012} which is used throughout this study. While DDA has proven to be effective and accurate in describing the optical properties of nanomaterials, it has not been previously applied in the context of SEIRA, to the best of our knowledge. Previous numerical investigations of SEIRA, mainly utilizing the Finite Difference Time Domain approach, have successfully replicated the Fano-like spectra but have not provided definitive conclusions regarding the origin of this phenomenon.

In Sec.III, we delve into the SEIRA response of a model molecular system, featuring a vibrational mode corresponding to the CO stretching mode, positioned within the hot spot created by two gold nanorods.


\section{Methods}

In DDA, the materials are modeled by a set of coupled polarizable dipoles excited by an external field, a monochromatic plane wave in our case. The formal derivation from Maxwell's equations and boundary conditions~\cite{yurkin_discrete_2007} reveals that the discretization of the bulk materials and the definition of the local polarizability play a crucial role in determining the simulation accuracy. For the gold nanorod, we derived the local polarizability from lattice dispersion relation, which is a correction of Clausius-Mossotti's formula~\cite{draine_discrete-dipole_1994}, from the experimental bulk refractive index~\cite{olmon_optical_2012}. The simulations have been performed with the DDSCAT code~\cite{draine_discrete-dipole_1994}

In our simulations, we consider nano-rods with a $30$ nm diameter and a length of $900$ nm, modeled by \num{199680} dipoles placed on a cubic grid with a periodicity of $1.875$ nm, ensuring the convergence (see also Ref~\cite{pelaez-fernandez_tuning_2023} for an investigation of the response of similar nano-object by DDA).

After the coupling with the external fields, the polarization of each dipole is computed and the local properties such as the electromagnetic near-field is determined. Far-field properties such as the absorption and scattering cross-sections are calculated by summing the contribution of each individual dipole and can be attributed to specific regions of the system.

To describe the response of the probed molecules in SEIRA, we take advantage of the versatility of the DDA by representing them as point dipoles with a polarizability associated with their vibrational modes described by a Drude-Lorentz model. The vibrational mode wavenumber is varied from $1400$ cm$^{-1}$ to $2000$  cm$^{-1}$, corresponding to the CO stretching modes in different chemical environments, to investigate the detuning between the LSPR and one of the molecular vibrational modes, with a damping parameter set at 16  cm$^{-1}$. Of course, in experimental systems, the vibrational frequency is fixed by the molecular composition and cannot be tuned. In the same way, if the plasmon resonance can be tuned by a modification of the geometry of the system, it cannot be modified dynamically. Our main goal in this study is to investigate the SEIRA effect with both the metallic nanoparticle and the molecules presenting an optical resonance at the same frequency or at nearby frequencies. For practical reasons, the metallic plasmonic system will be kept fixed in the following, while the molecular mode will be tuned. Note that the optical resonance of a nanorod can be modified thanks to LASER engraving~\cite{pelaez-fernandez_tuning_2023}. The molecules are modeled by 88 dipoles dispersed in the gap with a distance of at least $3.25$ nm with their closest neighbor which diminishes self-interacting phenomenon.

\begin{figure}
\begin{center}
\includegraphics[width=0.5 \textwidth] {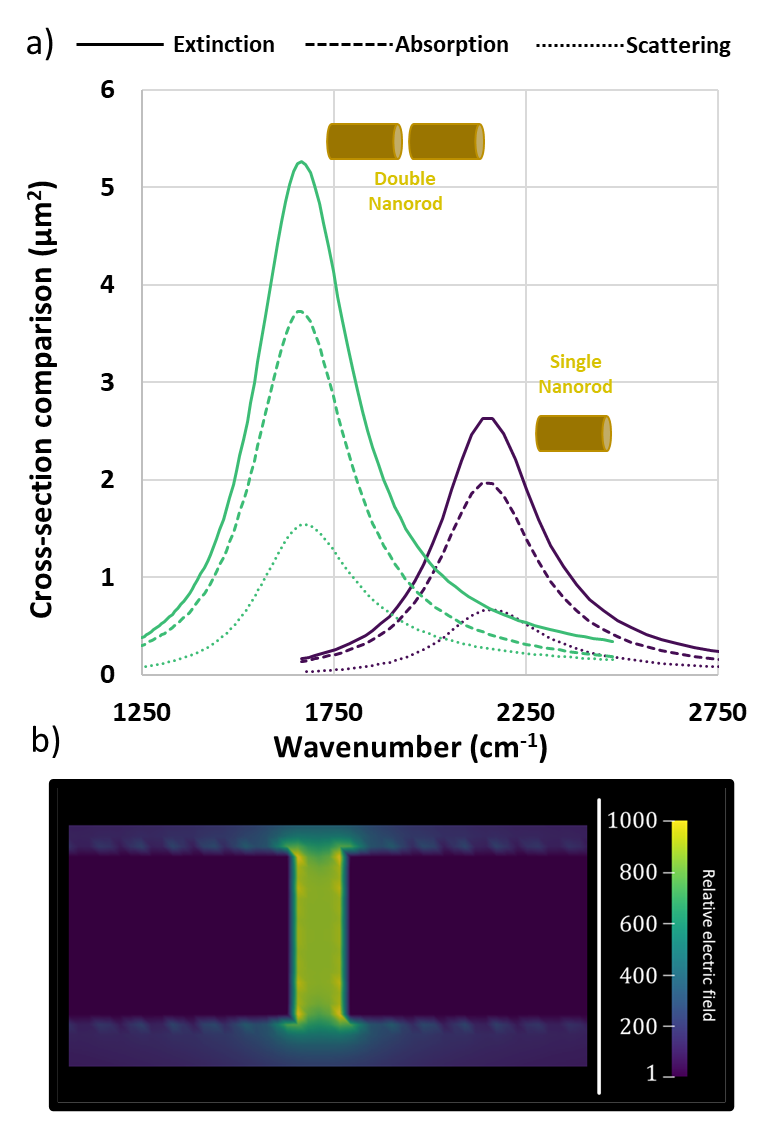}
\caption{\label{fig:rodRods} a) Extinction, absorption and scattering cross-section of the single nano-rod (right, purple) and the double nano-rod (left, green). b)  Electric field in a plane at the center of the gold nanorods, zoomed on the gap, at 1667 cm$\textsuperscript{-1}$. The enhanced electric field is normalized to the incident electric field amplitude.}
\end{center}
\end{figure}

\begin{figure}
\begin{center}
\includegraphics[width=0.5 \textwidth]{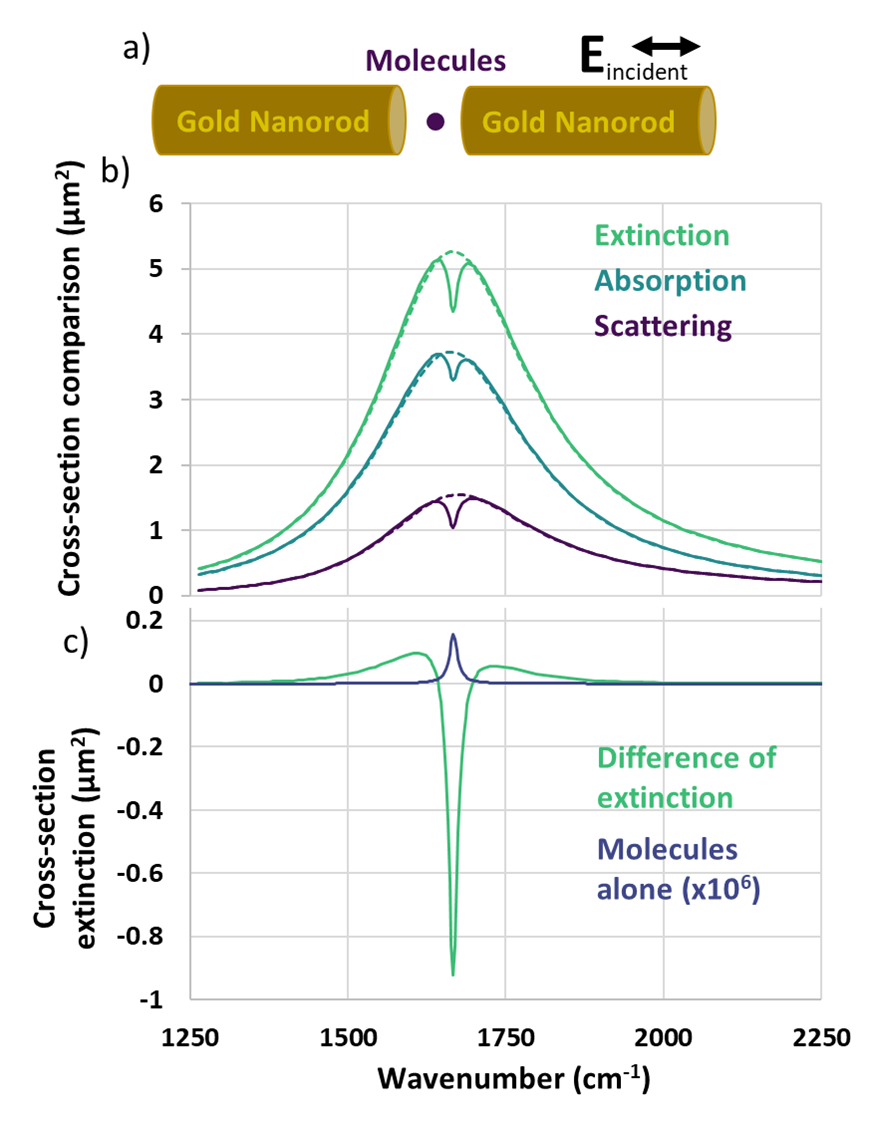}
\caption{\label{fig:Enh6000} a) Studied system with two gold nanorods separated by $9.375$ nm with molecules in between. The incident field is along the nanorods. b) Extinction, absorption and scattering cross-section of (a) with molecules (solid line) and without molecules (dotted line). c) Extinction cross-section difference between the situation with and without molecules (green line) compared with that of the molecules without gold nanorods (blue line). The latter has been multiplied by 10\textsuperscript{6} to ease the comparison.}
\end{center}
\end{figure}
  
\section{Results and Discussion}


The electromagnetic response of the gold nanorods in the infrared is due to the plasmon resonance that can be described, for an infinite nanorod, by the azimuthal number m and the wavevector along the nanorod axis~\cite{pelaez-fernandez_tuning_2023}. For finite nanorod of length $L$, the wavector is quantified ($k= n \pi /L $) and the m=1 mode is excited by light polarized perpendicular to the nanorod axis and the m=0 mode for parallel polarization~\cite{gomez-medina_mapping_2008}. The optical response of our gold nanorod is shown in \cref{fig:rodRods}a for a longitudinal excitation in the model wavenumber range of interest for this study. The resonance at $2150$  cm$^{-1}$ to $2000$ corresponds to the plasmon with the longest wavelength that can take place in the nanorod ($n=1$).

The coupling with a second nanorod modifies the optical response and creates a hot spot of intense local electromagnetic field, depending strongly on the separation distance. In \cref{fig:rodRods}a, a redshift to $1667$  cm$^{-1}$ is obtained for nanorods separated by $9.375$ nm, a distance small enough to get a strong hot spot but that can be described by our discretization grid. The corresponding near-field amplitude in the hot spot region at the resonance wavenumber is, in our case, enhanced by more than three orders of magnitude compared to the incident field (\cref{fig:rodRods}b).
 

We investigate the SEIRA response of the molecule lying  in the hot spot, as illustrated on \cref{fig:Enh6000}a. The wavenumber of the vibrational mode is chosen to be $1667$  cm$^{-1}$ to $2000$ corresponding to the maximum of the plasmon resonance of the coupled nanorods. 

The absorption, scattering and extinction cross-sections are shown with and without molecules in \cref{fig:Enh6000}b. We recover here the well-described decrease of the total cross-section in the presence of the molecule~\cite{yang_nanomaterial-based_2018,zvagelsky_plasmonic_2021,durmaz_multiple-band_2018,giordano_self-organized_2020,farmer_ultrasensitive_2016,li_graphene_2014,yan_tunable_2014,hwang_ultrasensitive_2021,langer_present_2020,ishikawa_cross-polarized_2017,neubrech_plasmonic_2013,neuman_importance_2015,neubrech_spatial_2014,dong_nanogapped_2017,neubrech_surface-enhanced_2017,wang_surface-enhanced_2019,wei_ultrasensitive_2019,hu_gas_2019,tanaka_nanostructure-enhanced_2022,mackin_plasmonic_2019,barbillon_latest_2022,cerjan_asymmetric_2016,abb_surface-enhanced_2014} that is observed for both absorption and scattering of light~\cite{neuman_importance_2015}. The difference of the extinction cross-sections (\cref{fig:Enh6000}c) evidences this transparency induced by the molecules more clearly but also displays the Fano-like profile with an increase of the extinction on the two sides of the main negative peak. 

It is hardly justified to speak about enhancement for SEIRA due to the very different shape of the response of the isolated molecule and of the SEIRA signal. This is even clearer for the Fano line shape when the resonance frequencies of the plasmonic system and of the molecules do not coincide (see later). We then prefer the term sensibility that we define by the difference between the extrema of the differential cross-section, divided by the maximum absorption cross-section of the isolated molecule, following Ref~\cite{baillieul_surface_2022}. In the example above, the sensibility is \num{6.54e5}. Experimentally, the sensibility is difficult to evaluate as the optical cross-section of the exact same plasmonic system without the molecules is not available and the number of molecules in the hot spot is not known.


\begin{figure}
\begin{center}
\includegraphics[width=0.5 \textwidth]{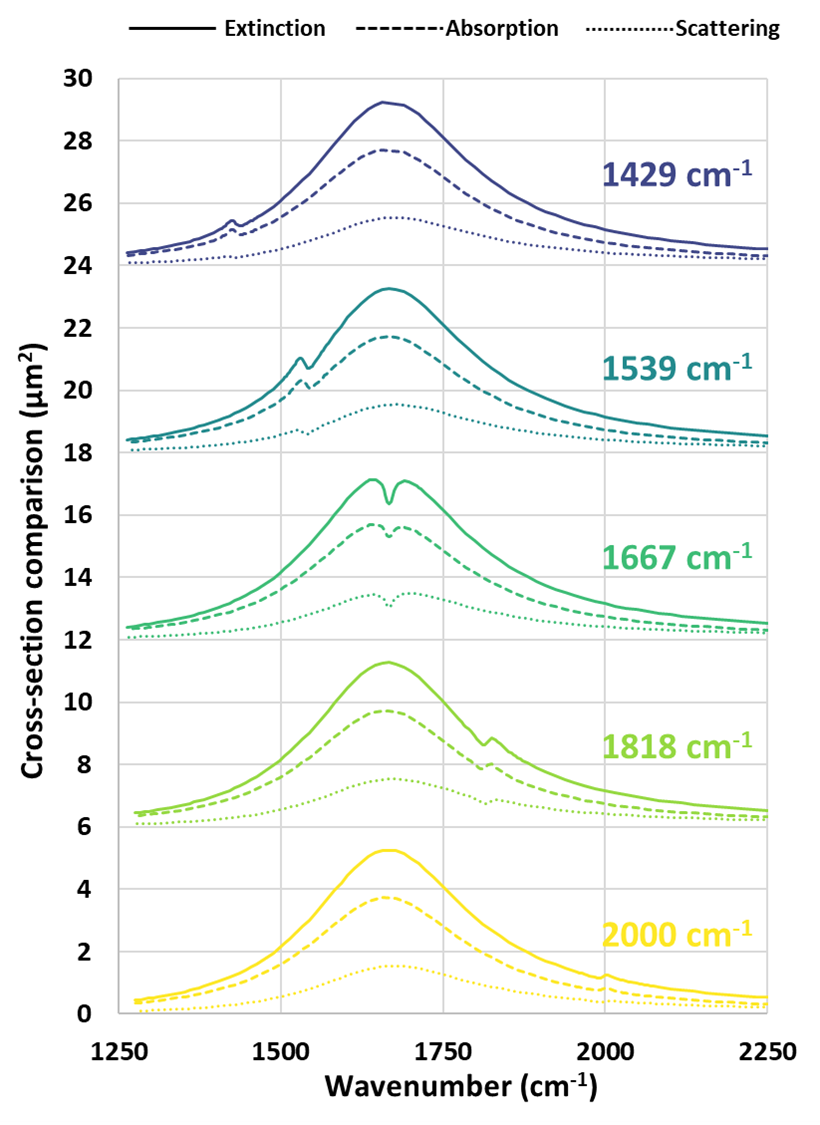}
\caption{\label{fig:allTheEnh} Extinction, absorption and scattering cross-section of \cref{fig:Enh6000}a with molecules. Each set of curves corresponds to a different model molecule whose vibrational  mode wavenumber is  indicated on the right, in the same color as the associated curves.}
\end{center}
\end{figure}

\begin{figure}
\begin{center}
\includegraphics[width=0.5 \textwidth]{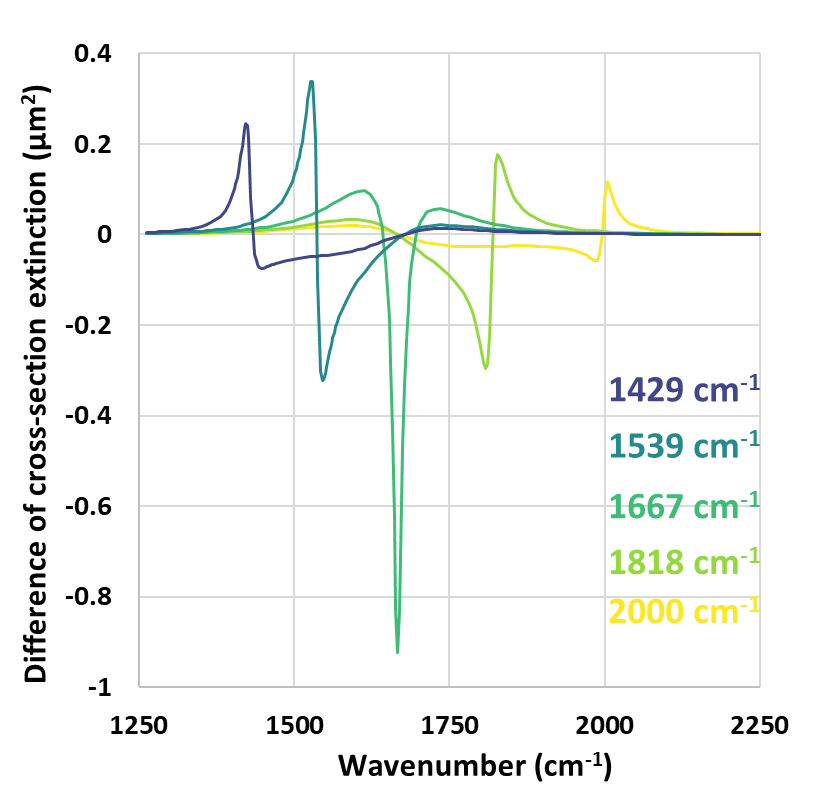}
\caption{\label{fig:onlyEnh} Extinction cross-section of \cref{fig:Enh6000}a with molecules subtracted by the extinction cross-section without molecules. Each curve is a model molecule with a different wavenumber for its vibrational mode. }
\end{center}
\end{figure}

The case where both resonances coincide is very specific and does not correspond to most of the experimental results. \cref{fig:allTheEnh} shows the extinction, absorption and scattering for five different vibrational wavenumbers below and above the plasmonic resonance. \cref{fig:onlyEnh} presents the Fano-like difference of cross-section, with a curve shape in agreement with previous studies~\cite{giordano_self-organized_2020,neuman_importance_2015,dong_nanogapped_2017,li_graphene_2014,yan_tunable_2014,zvagelsky_plasmonic_2021,cerjan_asymmetric_2016,langer_present_2020,neubrech_plasmonic_2013,neubrech_spatial_2014,neubrech_surface-enhanced_2017,wei_ultrasensitive_2019,tanaka_nanostructure-enhanced_2022}. The overall shape of the SEIRA signal depends strongly on the difference of the resonances between the particles and the molecules~\cite{tu_coupled_2010}, ranging from an induced transparency described above to, mainly, an increase of the extinction when the resonances do not coincide.


\begin{figure}
\begin{center}
\includegraphics[width=0.5 \textwidth]{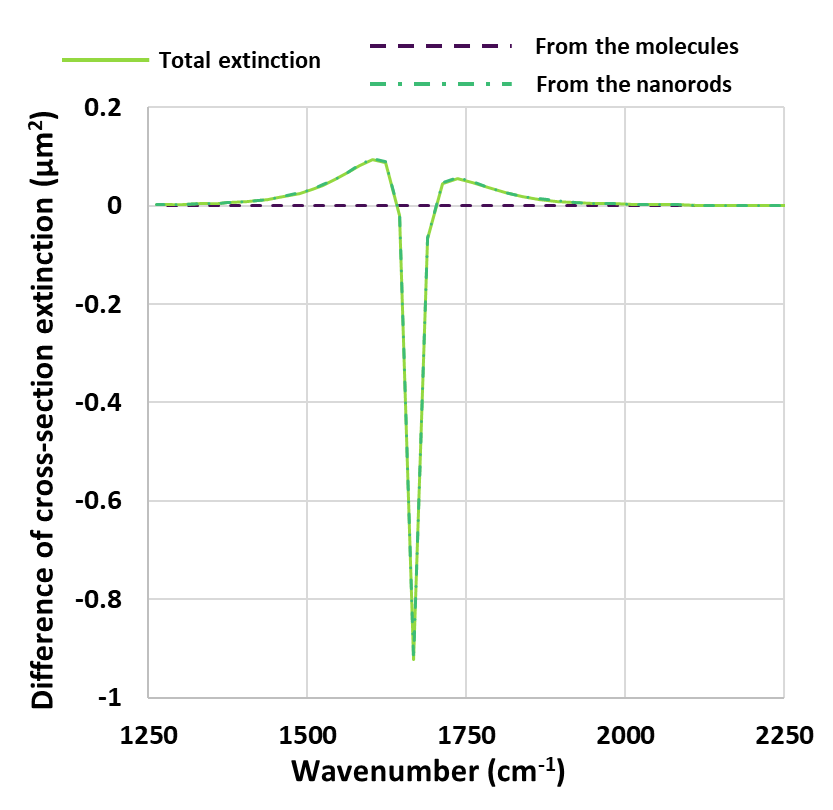}
\caption{\label{fig:rodOverMol} Extinction cross-section of \cref{fig:onlyEnh} for $1667$  cm$^{-1}$ to $2000$ (green line) with the contribution of the molecules (purple dash) and the contribution of the nano-rods (green dash and point) separated.}
\end{center}
\end{figure}

\begin{figure}
\begin{center}
\includegraphics[width=0.5 \textwidth]{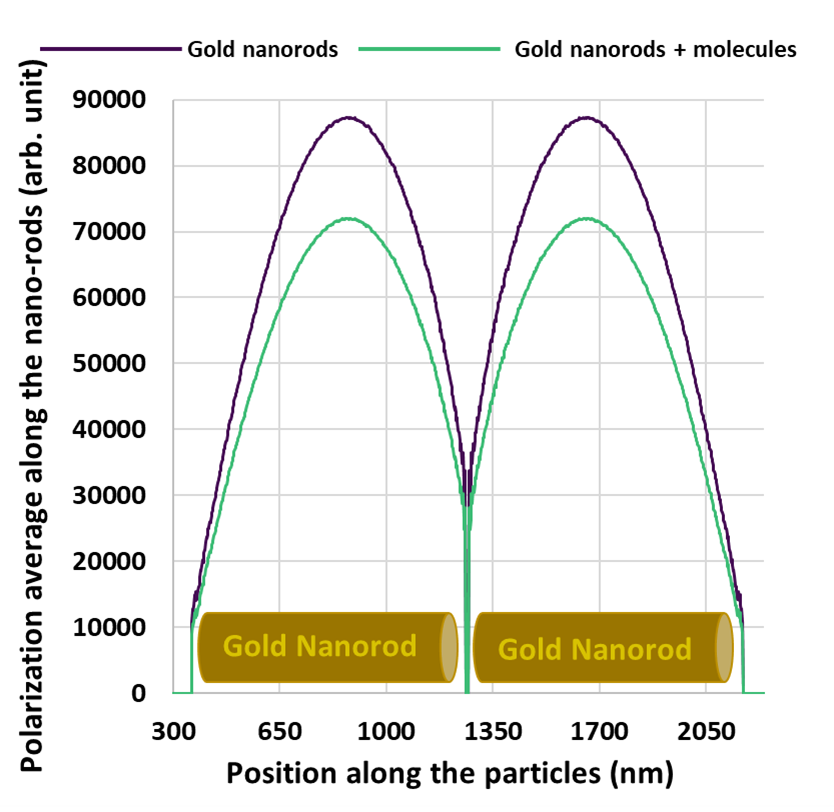}
\caption{\label{fig:popol} Polarization average along the nano-rods without molecules (purple top line) and with molecules (green bottom line).}
\end{center}
\end{figure}

In a DDA calculation, we can discriminate the spatial origin of the optical response as it is expressed as a sum over the local dipoles~\cite{draine_discrete-dipole_1994}. It turns out that the change of extinction originates almost entirely from the nanorods, as it is illustrated for the case of the strongest SEIRA response in \cref{fig:rodOverMol} (the same conclusion can be drawn for the other computed spectra). This conclusion is also supported by the analysis of the polarization profile of the nanorod along its axis with and without the molecules (\cref{fig:popol}). The polarization profile associated with the LSPR mode is very similar in both cases but with a global drop of the response. This behavior is similar to the one expected if a dielectric medium were added between the interacting nanorods which would make the LSPR resonance to be less pronounced, tying in with previous qualitative explanations~\cite{osawa_surface-enhanced_1993,yang_nanomaterial-based_2018}. 

Finally, it is interesting to note that the SEIRA response is observed in both the absorption and the scattering spectra (\cref{fig:allTheEnh} and Ref~\cite{neuman_importance_2015}). As there is no scattering from a single molecule (or from a point dipole in our model), the change in scattering comes entirely from the nanorods, going in the same direction as our previous conclusions. SEIRA is then associated with a modification of the LSPR resonance due to the presence of the molecules rather than to a direct contribution of the molecule to the spectra.


\section{Conclusion}

SEIRA is a very sensitive vibrational spectroscopy. Its modeling is not an easy task at it requires to describe at the same time, the plasmonic response of a metallic system and the vibrational response of a molecule. In this paper we have proposed to investigate SEIRA within a DDA approach, reproducing its main features such as the induced transparency and the Fano-like shape of the resonance for different mismatch of the resonance frequencies. 

We attribute the sensibility of SEIRA signature to a modification of the plasmon excitation due to the presence of the molecules rather than to a response of the molecules interacting with the local field. This is of prime importance for the design of new SEIRA samples as the more adapted plasmonic systems will be the ones that are more sensitive to a change of environment (the molecules) rather than the ones with the highest local-field enhancement. 

\ack
The calculations were performed on computers of the Consortium des Equipements de Calcul Intensif, including those of the Technological Platform of High-Performance Computing, supported by the FNRS-FRFC (Conventions No. 2.4.61707.F and 2.5020.11) and the University of Namur. The authors The authors acknowledge the funding by the Federation Wallonia-Brussels  through the ARC SURFASCOPE (UNamur-UCLouvain, Convention 19/24-102). V. Liégeois thanks the F.R.S.–FNRS for his Research Associate position.

\bibliographystyle{iopart-num}
\bibliography{biblio}

\end{document}